\documentclass{JINST}

\title{A BGO scintillating bolometer for $\gamma$ and $\alpha$ spectroscopy}

\author{L. Cardani$^a$\thanks{Corresponding author.}~ , S. Di Domizio$^b$ and L. Gironi$^c$\\
\llap{$^a$} Dipartimento di Fisica, Sapienza Universit\`{a} di Roma and Sezione INFN di Roma, Roma I--00185, Italy\\
\llap{$^b$} Sezione INFN di Genova,  Genova I--16146, Italy\\
\llap{$^c$} Dipartimento di Fisica, Universit\`{a} di Milano Bicocca and Sezione INFN di Milano Bicocca, Milano I--20126, Italy\\
E-mail: \email{laura.cardani@roma1.infn.it}}

\abstract{A 891 g BGO (Bi$_4$Ge$_3$O$_{12}$) scintillating bolometer has been tested at 10 mK 
in the underground Laboratori Nazionali del Gran Sasso (Italy).
The discrimination capability, the radio-purity of the compound and the main features of the crystal have been studied in order
to demonstrate the excellent performances obtained by operating a scintillating bolometer in the field of $\gamma$ and $\alpha$ spectroscopy.
The sensitivity of this detector in the study of extremely low surface contaminations has been investigated.}


\begin{document}

\section{Introduction}

The neutrinoless double beta decay (0$\nu$DBD) has a widely recognized importance in modern physics,
as its direct observation would demonstrate the Majorana nature of neutrino and set a value on the absolute mass of this particle.

Cuoricino, designed for the purpose of studying the 0$\nu$DBD of $^{130}$Te, was operated from 2003 to 2008 in the Laboratori Nazionali del Gran Sasso (Italy)
and is still one of the most sensitive experiments ever realized in this field \cite{CUORICINO}. Cuoricino was based on the bolometric technique and consisted of 62 TeO$_2$ crystals (corresponding to 11.3 kg of $^{130}$Te) operated at a constant temperature of $\sim$10 mK.
Thanks to the excellent performances provided by this detector in terms of efficiency and energy resolution,
Cuoricino could set an upper limit on the neutrino Majorana mass of 300 -- 710 meV, depending on the adopted nuclear matrix element evaluation.

The important physics results provided by this experiment lead to its upgrade, CUORE,
that will improve the  discovery potential on the effective Majorana neutrino mass to 41 -- 95 meV \cite{CUORE, sensitivity}.
The achievement of this goal is subject to the suppression of the background counting rate in the energy region of the $^{130}$Te 0$\nu$DBD ($\sim$2527 keV) down to 10$^{-2}$ counts/keV/kg/y, which means about an order of magnitude less with respect to Cuoricino.

Such a low background can not be reached only by locating the experiment in deep underground laboratories and by equipping it with proper shields. The Cuoricino experience, indeed, showed that the most important sources of background are the radioactive contaminations of the materials located close to the detector.
For this reason, the materials of CUORE must have a bulk contamination in $^{238}$U e $^{232}$Th (or their daughters) lower than a few pg/g,
as well as surface contaminations lower than a few nBq/cm$^2$.
Unfortunately, due to the low level of the contaminations, these sources of background are very difficult to identify and reduce.

Bulk contaminations of the materials can be studied with a good sensitivity through conventional techniques, i.e. high purity germanium (HpGe), inductively coupled plasma mass spectrometry (ICP-MS) and neutron activation.

The situation is much more complicated for the material surface contaminations.
Since ``standard'' techniques (i.e. Si Surface Barrier Detectors) can not achieve a sufficient sensitivity,  the only way to access these extremely low contaminations is the bolometric technique.
Bolometers, indeed, have no dead layer and can reach energy resolutions orders of magnitude better than standard Silicon Surface Barrier Detectors.

Unfortunately, pure bolometers can not identify the nature of the particles that interact in the detector and so the characterization of the background sources remains a very difficult task.

In this paper we demonstrate how this problem can be overcome by means of a BGO scintillating bolometer.

\section{Scintillating Bolometers}
Bolometers are calorimeters in which the energy released in the absorber by an interacting particle is converted into phonons and measured via a temperature variation \cite{bolometers}. 
An ideal instantaneous deposition of energy E in the absorber gives rise to a temperature variation $\Delta$T $\propto$ E/C, where C is the thermal capacitance of the bolometer itself. 
A small capacitance is therefore needed to obtain sizable $\Delta$T. For this reason, bolometers are usually made of dielectric and diamagnetic crystals and are operated at cryogenic temperature ($\approx$10 mK).

The temperature increase is measured through dedicated sensors, whose resistance shows a steep dependence on the temperature. For Neutron Transmutation Doped (NTD) Ge thermistors \cite{NTD}, the resistance variation is than converted into a readable voltage signal by biasing the thermistors with a constant current.

If the bolometer is a scintillating crystal, the heat signal can be combined with the light signal. The composite device shows many advantages: the bolometric technique provides an excellent energy resolution and a high efficiency, while the simultaneous read-out of the light allows to identify the interacting particles (n, $\alpha$, $\beta/\gamma/\mu$) thanks to their different Light Yield (LY). Indeed, the bolometers are  very slow detectors (pulse time development of the order of hundreds of milliseconds) and do not allow to recognize the interacting particle by the shape of the thermal signal except in some cases in which the energy release in the crystals produces states with very long decay times \cite{bol_pulse_shape}.\\
Examples of other interesting features of scintillating bolometers can be found in \cite{nsbolometriscintillanti}.\\
Our purpose is to demonstrate that an accurate characterization of the surface contaminations of a material can be obtained in a simple way by facing the material itself to a scintillating bolometer.
The precise measurement of the particle energy, as well as the tagging of the particle nature, allows for the identification of the single contaminants. At the same time, the large active surface and the absence of dead layers provide a very high sensitivity.

In this paper we present the results of a measurement performed with a 5$\times$5$\times$5 cm$^3$ BGO  scintillating crystal (891 g).
The choice of this crystal is due to many interesting features: first of all, the radio-purity of the compound and the large light yield. Then, the high atomic number of BGO and the large dimensions of our sample, which make this detector very appealing also for $\gamma$ spectroscopy.

\section{Detector}
\begin{figure}[!htb]
\centering
\includegraphics[width=.6\textwidth]{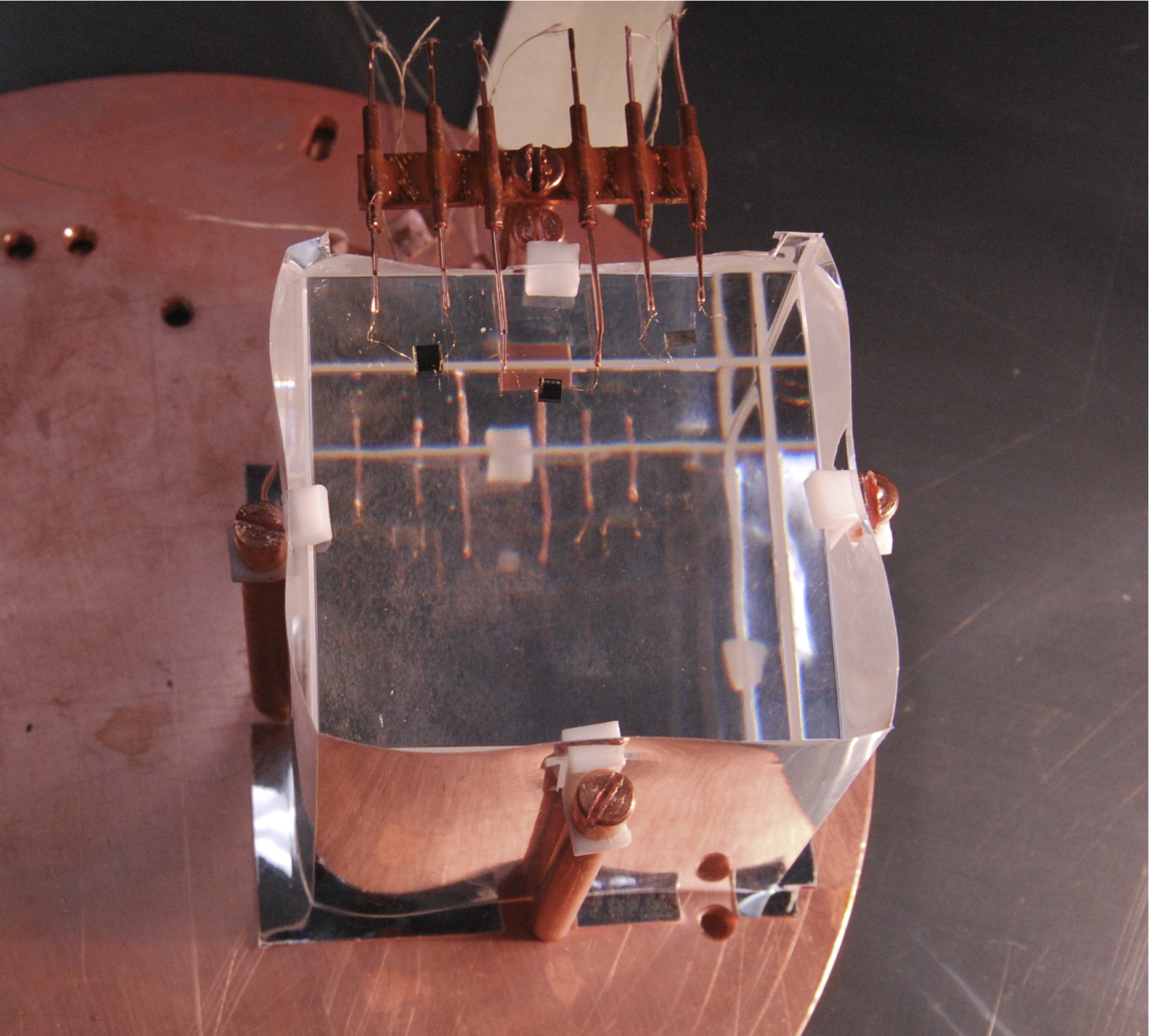}
\caption{The 5$\times$5$\times$5 cm$^3$ BGO crystal surrounded by the reflecting foil.}
\label{fig: BGO setup1}
\end{figure}

The detector we present is a BGO crystal whose surfaces are polished at optical grade. The crystal is held by means of L-shaped Teflon pieces (PTFE) fixed to copper columns, and it is surrounded by a reflecting foil (3M VM2002) in order to increase the light collection efficiency (see figure \ref{fig: BGO setup1}).

The emitted light is measured by a second bolometer. This Light Detector (LD) is realized with a 36 mm diameter and 0.3 mm thick pure germanium disk covered with a 600 $\mathring{A}$ thick layer of SiO$_2$ in order to ensure a good light absorption.

The composite detector is mounted in the Oxford 200 $^3$He/$^4$He dilution refrigerator located in the Laboratori Nazionali del Gran Sasso ($\approx$3650 m w.e.). The details of the cryogenic facility can be found in \cite{Hall C}.

The signals produced by the thermistors are read via a cold pre-amplifier stage located inside the cryostat.
A second stage of amplification, as well as the front end, are located on the top of the cryostat at room temperature. The signals are filtered by means of an anti-aliasing Bessel filter
with cut frequencies of 63 Hz for the BGO signals and 120 Hz for the Light signals  \cite{Bessel}. The entire waveform is digitized and recorded with a sampling frequency of 2 kHz.\\ 
The trigger of the BGO and of the LD is software generated. The LD is acquired in coincidence with the heat channel in order to reject most of the signals that are not due to a BGO scintillation event.

The off-line analysis allows to determine the pulse amplitude, as well as many pulse parameters, by means of the optimum filter technique \cite{Optimum filter, GironiZnMoO4}.\\

\section{Results}

The BGO crystal cooled very slowly.
A similar behaviour was noted also in a previous measurement with 2$\times$2$\times$2 cm$^3$ BGO crystals.
This behaviour can be ascribed to some excess heat capacity that decouples at low temperatures. In spite of the cooling problem, it was possible to perform an accurate measurement with this crystal.

\begin{figure}[!htb]
\centering
\includegraphics[width=.6\textwidth]{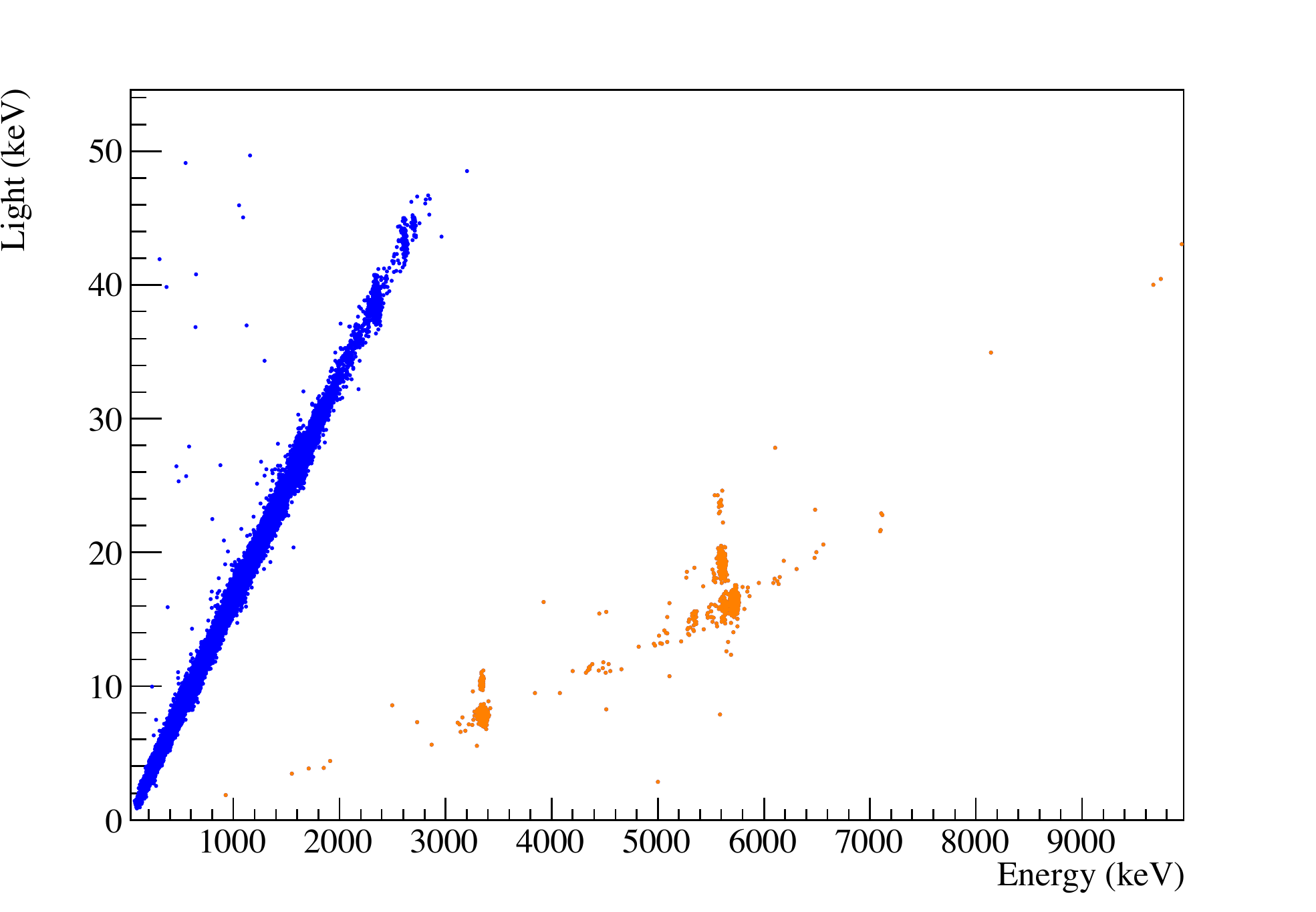}
\caption{Light vs heat scatter plot of the 5$\times$5$\times$5 cm$^3$ BGO crystal for a $\approx$455 hours background measurement. In blue $\beta/\gamma$ events, in orange $\alpha$ events.}
\label{fig:scatterBGO 5x5x5}
\end{figure}

\begin{figure}[!htb]
\centering
\includegraphics[width=.6\textwidth]{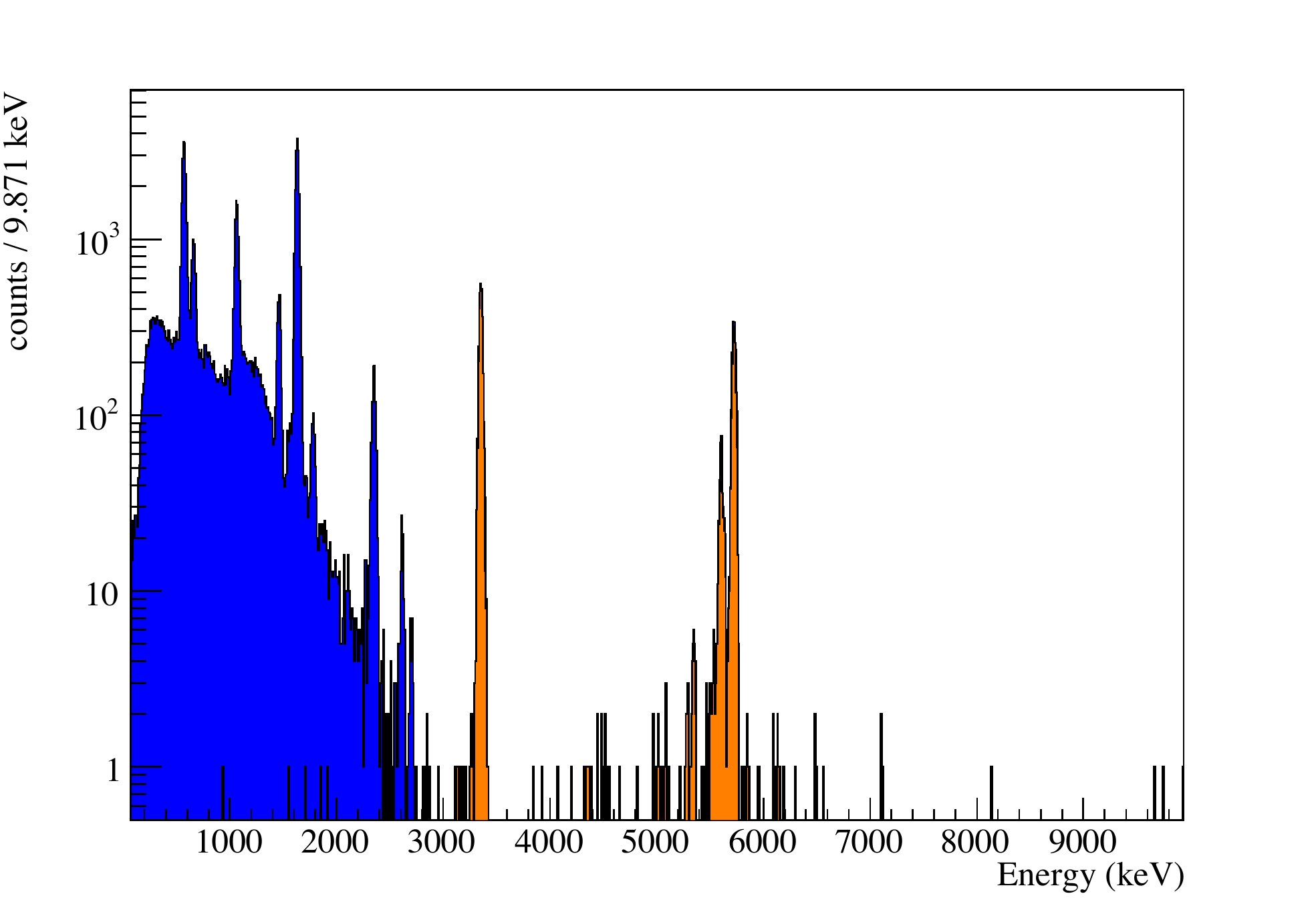}
\caption{Heat Spectrum of a 5$\times$5$\times$5 cm$^3$ BGO crystal for a $\approx$455 hours background measurement. 
Blue: $\beta/\gamma$ events. Orange: $\alpha$ events.}
\label{fig:spectrumBGO 5x5x5}
\end{figure}

The light vs heat scatter plot obtained in $\approx$455 hours of measurement is reported in figure \ref{fig:scatterBGO 5x5x5}.
The excellent discrimination between $\alpha$ and $\beta/\gamma$ events can be observed.\\
The heat spectrum (reported in figure \ref{fig:spectrumBGO 5x5x5}) is calibrated on the $^{208}$Tl line and on the $\gamma$ peaks of $^{207}$Bi which is produced by proton interactions on $^{206}$Pb \cite{207Bi_N}.  It is probably mainly originated by the interaction of cosmic ray protons since the BGO crystal was produced in China (SICCAS) and delivered by plane.

\begin{figure}[!htb]
\centering
\includegraphics[width=.6\textwidth]{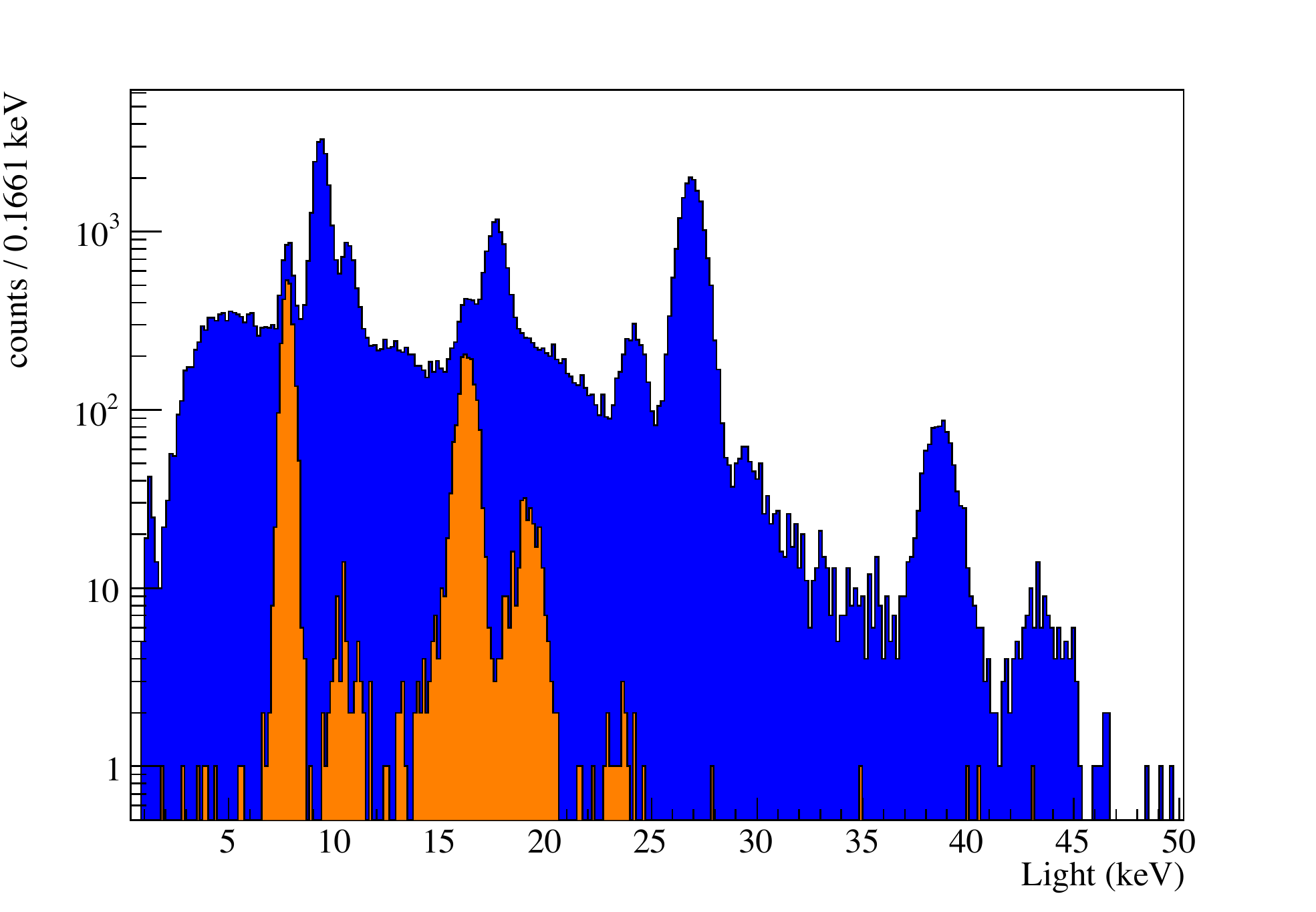}
\caption{Light Spectrum of a 5$\times$5$\times$5 cm$^3$ BGO crystal for a $\approx$455 hours background measurement. 
Blue: total spectrum. Orange: $\alpha$ events.}
\label{fig:light_spectrumBGO 5x5x5}
\end{figure}

The light spectrum (figure \ref{fig:light_spectrumBGO 5x5x5}) is calibrated by means of a $^{55}$Fe source faced to the LD. 
The calibration is performed with a simple linear function, by attributing the nominal energy of the $^{55}$Fe X-rays (5.89 keV and 6.49 keV) to the identified peaks.\\
The radio-purity of the crystal has been studied by investigating its internal contaminations.

In the $\alpha$ region several peaks can be identified. These are due to internal contaminations of $^{209}$Bi, $^{210m}$Bi and $^{210}$Po.

$^{210m}$Bi is produced via the $^{209}$Bi(n,$\gamma$)$^{210m}$Bi reaction and accumulates in the crystal due to its long half-life (T$_{1/2}$=3$\times$10$^6$ y) \cite{210Bi}. $^{210m}$Bi alpha-decays on $^{206}$Tl with a Q-value of $\approx$5.3 MeV. The decay can be on the fundamental level, giving rise to the lowest spot at 5.3 MeV (see figure \ref{fig:scatterBGO 5x5x5}), or on the excited state of  $^{206}$Tl with the prompt emission of $\gamma$ rays of $\sim$300 and $\sim$640 keV. Since the probability that these $\gamma$ rays release all their energy in the BGO crystal itself is high ($>$54\%), these mixed $\alpha$+$\gamma$ events give rise to other two spots that lie between the $\alpha$ and the $\beta$/$\gamma$ band.

$^{209}$Bi (Q$=$3137 keV) was supposed to be the heaviest stable isotope, until P. de Marcillac \emph{et. al} detected its decay \cite{Pierre}.
Thanks to the simultaneous observation of the ground and excited state decay of this isotope,
our measurement provides the ultimate test on the rare decay of $^{209}$Bi \cite{209Bi excited}.
       
Finally, the high number of counts due to the decay of $^{210}$Po (Q$=$5407 keV) is probably due to the activation of $^{209}$Bi by the $^{209}$Bi(n,$\gamma$)$^{210}$Bi reaction. $^{210}$Bi subsequently gives origin to $^{210}$Po through $\beta$-decay (T$_{1/2}$ $\approx$ 5 days).

The full understanding of these contaminants allows for a detailed analysis of the resolution and of the light emission of the crystal.

The FWHM energy resolution of the heat channel is (37.5 $\pm$ 0.5) keV, with no dependence on the energy of the peak. 
This value is obtained using the energy spectrum of the entire measurement. 
However, better results can be obtained fitting only the spectrum of the last days of measurement,
as the cooling of the crystal determined an increase in the signal to noise ratio and, therefore, an improvement in the energy resolution.
In a couple of weeks of measurement, indeed, we observed an in improvement of the resolution of 51\%.

The performance of the light detector is excellent. The energy FWHM resolution of the light signals ranges from 0.78 keV to 1.5 keV (the first value calculated on the 570 keV peak of $^{207}$Bi, the second one on the 2615 keV line of $^{208}$Tl).

The analysis of the light emission of the BGO crystal demonstrates the discrimination capability of this detector.

The LY of the $\gamma$ peaks is evaluated on $^{40}$K, $^{208}$Tl and each line of $^{207}$Bi and no dependence on the energy is observed. The average value is LY($\beta/\gamma$) = (16.61 $\pm$ 0.02) keV/MeV.

On the contrary, the LY of $\alpha$ particles increases with the energy: 
we measured LY = (2.482 $\pm$  0.002) keV/MeV at 3137 keV ($^{209}$Bi line) and LY = (3.011 $\pm$  0.003) keV/MeV at 5400 keV ($^{210}$Po line),
corresponding to a quenching factor for $\alpha$ particles of  0.1494 $\pm$  0.0002 and 0.1813 $\pm$  0.0002, respectively.
This behavior, which is a consequence of the energy dependence of the stopping power (Birks law \cite{Birks}), was already observed in other crystals  \cite{CdWO4}.

The low values of the quenching factor for $\alpha$'s, together with the narrow bands allow to discriminate very well $\alpha$ and $\beta/\gamma$ events. 

In addition, except for the $^{207}$Bi, $^{209}$Bi and $^{210m}$Bi isotopes, no evidence of internal radioactive contamination of the crystal is observed. In particular natural radioactive contaminations have an activity lower than 0.14 pg/g in $^{238}$U and 0.41 pg/g in $^{232}$Th at 90$\%$ C.L.
Most of the detected contaminations can be easily reduced by avoiding the shipping of the crystals by plane;
in this way the cosmic ray activation can be strongly suppressed.

The measurement presented in this paper allowed to perform a preliminary study of the surface contaminations. The analysis was focused on the $\alpha$ band in the 2-3 MeV region and in the 3.5-4 MeV region (the division is due to the presence of the $^{209}$Bi peak). The study of surface contaminations can be of particular interest for 0$\nu$DBD experiments, whose goal is the achievement of a low background counting rate in this energy region. The results are reported in Table \ref{tab:BGOBack}. The evaluation of the surface activity is obtained assuming conservatively that all the alpha events in the selected energy regions are due to surface contaminations of materials facing the detector.
The background obtained in the $\beta/\gamma$ region with E$>$3 MeV is also reported. The measured value also includes events due to pile-up of $\beta/\gamma$ due mainly to the high internal contamination of $^{207}$Bi. As mentioned above this rate can be reduced simply by avoiding the shipping of the crystals by plane.

The $\beta/\gamma$ region with E$>$3 MeV is very interesting especially for experiments aiming to study the 0$\nu$DBD of isotopes with large Q--value. Indeed, above this line the environmental background due to natural $\gamma$'s  decrease abruptly.

\begin{table}
\centering
\begin{tabular}{lccc}
\hline
Band  & Energy  &  \multicolumn{2}{c}{Background}      \\        
     & [MeV]   & [c/keV/kg/y]  & [nBq/cm$^2$/MeV]  \\   
     \hline
$\beta$ & 3.0 - 4.0     & 0.01 - 0.07   & --- \\    
$\alpha$ & 2.0 - 3.0    & 0.03 - 0.14  & 6.32 - 30.45 \\  
$\alpha$ & 3.5 - 4.0    & 0.04 - 0.22  & 8.50 - 48.84 \\
\hline
\end{tabular}
\caption{Background obtained with the 5$\times$5$\times$5 cm$^3$ BGO crystal in three different regions. 
The values are evaluated with Feldman-Cousins at 68.27\% C.L. The results  were obtained by assuming that all the $\alpha$ events were due to surface contaminations.}

\label{tab:BGOBack}
\end{table}
The very low counting rate obtained in the energy region of interest is rather surprising if one considers that we did not apply any cleaning procedure on the crystal and on the reflecting foil that was surrounding it.\\

\begin{figure}[!htb]
\centering
\includegraphics[width=.6\textwidth]{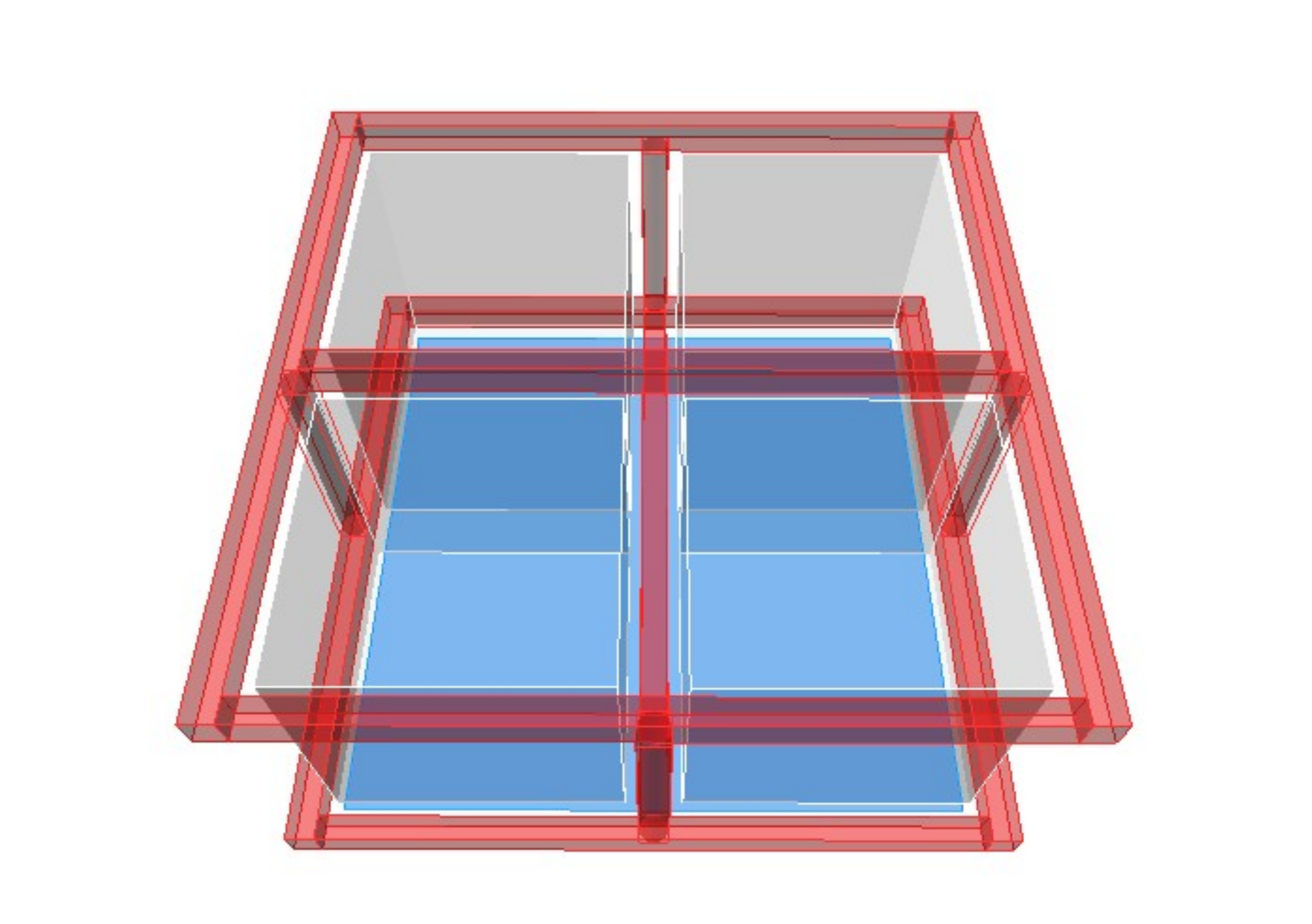}
\caption{Simulated array of four 5$\times$5$\times$5 cm$^3$ BGO crystals faced to a slab of material positioned under the crystals.}
\label{fig:BGO_Array}
\end{figure}

Since this technique will be applied to the study of surface contaminations, we performed Monte Carlo  simulations (GEANT-4) to understand the efficiency of this detector and then its sensitivity. 
The simulations were performed with an array of four 5$\times$5$\times$5 cm$^3$ BGO crystals faced to a slab of material to be analyzed (figure \ref{fig:BGO_Array}). 
The goal of these simulations is to study the geometric efficiency ($\epsilon$), i.e. the ratio between the number of alpha particles that hit the detectors and the number of generated particles on the surface facing the detectors. For this purpose, the simulations were performed generating $\alpha$ decays on the surface of the material (no depth).

In addition, different simulations were performed with $\alpha$ particles with energy between 3 and 6 MeV to study the dependence of the geometric efficiency on the particle energy. 
The simulations confirmed our expectations, showing that the energy dependence is less than 1$\%$ (limit given by the simulation statistics).

The geometric efficiency for the detection of surface contaminations of a circular/square slab with a diameter/side of 12.4/11 cm faced to the BGO crystal array at a distance of 5 mm was evaluated to be $\epsilon$ $\sim$35\%.
For a measurements of 1 month with a sample of 121 cm$^2$ faced to the crystals this allows to achieve sensitivity of 21 nBq/cm$^2$.
The obtained sensitivity decreases with increasing depth of the contamination because of the worsening of the detection efficiency.\\

In order to compare the sensitivity that could be reached with BGO crystals with respect to a standard $\alpha$ detector (i.e. Si surface barrier detector) and pure thermal bolometers (TeO$_2$ bolometers), several simulations were performed. 
The geometry of the simulation of the TeO$_2$ crystals was the same used for the simulation of the BGO.
For what concerns surface barrier detector, we simulated a 4 cm diameter Si disk \cite{CANBERRA} faced to the same slab of material to be analyzed.

\begin{figure}[!htb]
\centering
\includegraphics[width=.6\textwidth]{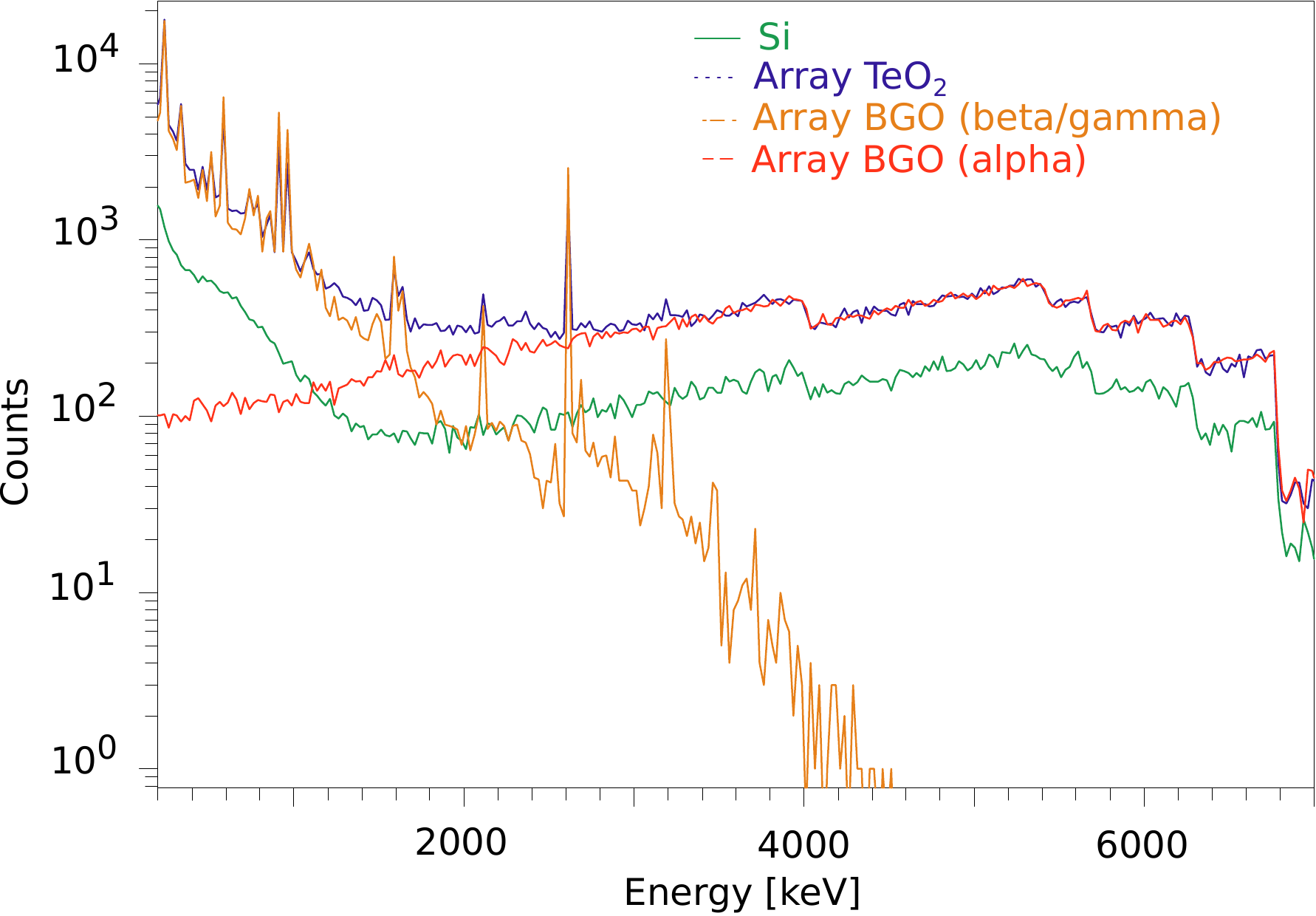}
\includegraphics[width=.6\textwidth]{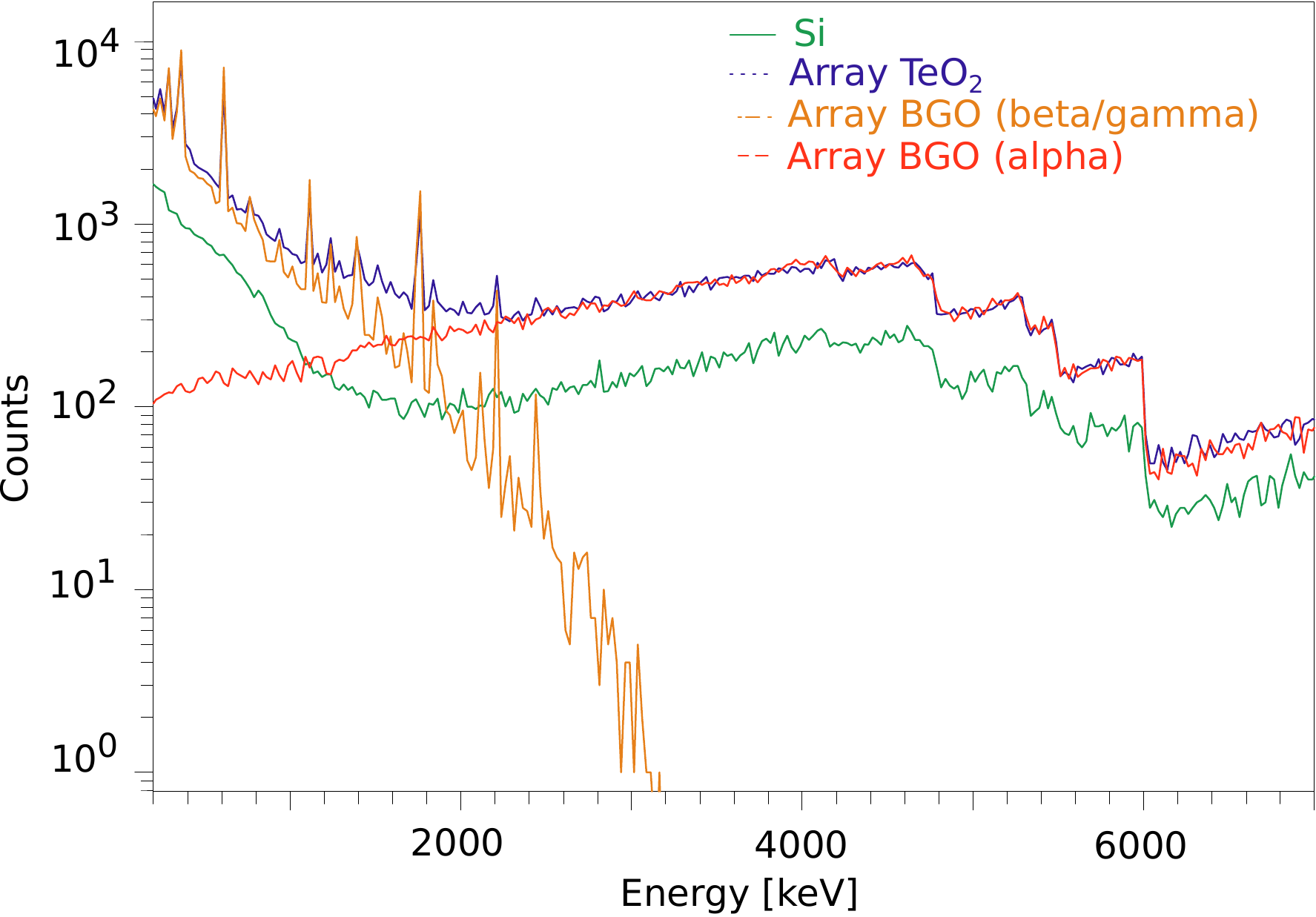}
\caption{Simulations of energy spectra that can be obtained with a surface contamination in $^{232}$Th (on the top) and $^{238}$U (in the bottom) with an exponential density profile and a depth $\lambda$ of 5 micron by a Si surface barrier detector (green), an array of four TeO$_2$ crystals (blue) and an array of four BGO crystals. For the last, thanks to the possibility to identify the interacting particle, the beta/gamma region (orange) and the alpha region (red) are reported separately.}
\label{fig:Simulazione_5micron}
\end{figure}

The results are reported in figure \ref{fig:Simulazione_5micron} for surface contaminations in $^{232}$Th and  $^{238}$U with an exponential density profile and a depth $\lambda$ of 5 micron (i.e. the one used for Cuoricino background reconstruction \cite{ExpProfile}).

As can be seen in the figures, the main advantage of the BGO detector is the possibility to perform separately  $\alpha$ and $\beta/\gamma$ spectroscopy.
This feature allows to study with a very high sensitivity low energy $\alpha$'s, that in the case of TeO$_2$ and Si detector would be overwhelmed by the $\gamma$ natural radioactivity.
At the same time, with the BGO detector it is possible to access  rare $\gamma$'s and  $\beta$ events in the energy region dominated by the $\alpha$ background (above 2.6 MeV).

Another important advantage of the BGO detector is the very low intrinsic background,
which is  about two orders of magnitude lower with respect to the count rate of a standard Si detector (0.05 counts/h/cm$^2$ between 3 and 8 MeV \cite{CANBERRA}).
This leads to a much better signal to noise ratio and, therefore, to a much larger sensitivity on surface contaminations.

Finally it must be stressed that, unlike Si detectors, bolometers can easily achieve large active surfaces and do not have dead layers. 
Moreover, the excellent energy resolution provided by bolometers ($\sim$5 keV FWHM energy resolution with a TeO$_2$ bolometer \cite{CCVR}
with respect to 25--30 keV FWHM resolution of a Si detector) makes these detectors particularly appealing for high sensitivity spectroscopy.

\section{Conclusion}
For the first time a large BGO crystal (891 g) was measured at cryogenic temperature, giving excellent results in terms of discrimination capability and radio-purity.

We demonstrated that this crystal is a perfect candidate to perform $\gamma$ and $\alpha$ spectroscopy
with a very high sensitivity.

Looking at these results, we are planning a measurement with an array of 4 BGO crystals (5$\times$5$\times$5 cm$^3$). The array will allow for a coincidence analysis and, therefore, for a precise determination of the surface contamination of the crystals. In addition the reflecting foil will be removed. In this way it will be possible to measure the surface contaminations of the materials close to the detector with a sensitivity of $\approx$ nBq/cm$^2$ in few weeks.

\end{document}